\documentclass[runningheads]{llncs}
\usepackage[hidelinks]{hyperref}
\usepackage[font=scriptsize]{caption}
\usepackage{graphicx}
\usepackage{amsfonts}
\usepackage{amsmath}
\usepackage{bbm}
\usepackage{diagbox}
\usepackage{comment}
\usepackage{listings}
\usepackage{tikz}
\usepackage{enumitem}

\usetikzlibrary{tikzmark}

\newcommand\restr[2]{{
  \left.\kern-\nulldelimiterspace 
  #1 
  \vphantom{\big|} 
  \right|_{#2} 
}}

\begin{document}
\title{Filtering and Sampling Object-Centric Event Logs}
\titlerunning{Filtering and Sampling OCELs}
%
\author{
Alessandro Berti\inst{1,2}
}
\authorrunning{A. Berti et al.}
%
\institute{Process and Data Science Group @ RWTH Aachen, Aachen, Germany
\email{\{a.berti, wvdaalst\}@pads.rwth-aachen.de }\\
\and
Fraunhofer Institute of Technology (FIT), Sankt Augustin, Germany\\}
\maketitle              
\begin{abstract}
The scalability of process mining techniques is one of the main challenges to tackling the massive amount of event data produced
every day in enterprise information systems. To this purpose, filtering and sampling techniques are proposed to keep a
subset of the behavior of the original log and make the application of process mining techniques feasible.
While techniques for filtering/sampling traditional event logs have been already proposed, filtering/sampling object-centric
event logs is more challenging as the number of factors (events, objects, object types) to consider is significantly higher.
This paper provides some techniques to filter/sample object-centric event logs.
\end{abstract}


\section{Introduction}
\label{sec:introduction}

Process mining is a branch of data science that exploits the event data contained in the information systems supporting mainstream processes (ERP, CRM)
to infer process-related information.
The primary assumption to apply process mining is the association of a \emph{case notion} to the events.
A case notion groups the events that belong to the same process execution.
In a ticket management system, the identifier of the ticket is used to group all the events (creation, problem resolution, closure) related
to the same ticket. However, in most of the other processes, the identification of a reasonable case notion is difficult. For example, in an Order-to-Cash process,
a sales order might be associated with many deliveries.
If the identifier of the sales order is chosen as a case notion, several instances of the activities done for different deliveries (create delivery, delivery successful) appear in the same case,
and the {\it divergence} problem occurs. Conversely, if the identifier of the delivery is chosen as a case notion, we would need to record the same events related to the order (create order, pay order)
in the same case, and the {\it convergence} problem occurs.
\begin{table*}[!t]
\caption{Tabular representation of an object-centric event log.}
\resizebox{\textwidth}{!}{ 
\begin{tabular}{|lcc|ccccc|}
\hline
{\bf Event ID} & {\bf Event Activity} & {\bf Event Timestamp} & {\bf Orders} & {\bf Items} & {\bf Deliveries} & {\bf Weight Classes} & {\bf Goods Issues} \\
\hline
e1 & Create Order & 2007-04-01 07:29 & ['o1'] & ['i1', 'i2'] & ~ & ['Normal'] & ~ \\
e2 & Pick Item & 2007-04-01 15:36 & ['o1'] & ['i1'] & ~ & ['Normal'] & ~ \\
e3 & Pack Item & 2007-04-01 22:01 & ~ & ['i1'] & ['d1'] & ['Normal'] & ~ \\
e4 & Pick Item & 2007-04-01 22:15 & ['o1'] & ['i2'] & ~ & ['Normal'] & ~ \\
e5 & Pack Item & 2007-04-01 23:19 & ~ & ['i2'] & ['d1'] & ['Normal'] & ~ \\
e6 & Deliver Successful & 2007-04-02 10:01 & ~ & ~ & ['d1'] & ['Normal'] & ~ \\
e7 & Pay Order & 2007-04-02 11:15 & ['o1'] & ~  & ~ & ['Normal'] & ~ \\
e8 & Create Order & 2007-04-02 12:15 & ['o2'] & ['i4', 'i5', 'i6'] & ~ & ['Normal'] & ~ \\
e9 & Pick Item & 2007-04-03 02:19 & ['o2'] & ['i4'] & ~ & ['Normal'] & ~ \\
e10 & Pack Item & 2007-04-03 03:17 & ~ & ['i4'] & ['d2'] & ['Normal'] & ~ \\
e11 & Pick Item & 2007-04-03 05:02 & ['o2'] & ['i5'] & ~ & ['Normal'] & ~ \\
e12 & Record Goods Issue & 2007-04-03 05:07 & ~ & ['i5'] & ~ & ['Normal'] & ['r1'] \\
e13 & Pick Item & 2007-04-03 05:15 & ['o2'] & ['i5'] & ~ & ['Normal'] & ~ \\
e14 & Pack Item & 2007-04-03 05:21 & ~ & ['i5'] & ['d3'] & ['Normal'] & ~ \\
e15 & Pick Item & 2007-04-03 05:36 & ['o2'] & ['i6'] & ~ & ['Normal'] & ~ \\
e16 & Pack Item & 2007-04-03 05:37 & ~ & ['i6'] & ['d3'] & ['Normal'] & ~ \\
e17 & Delivery Successful & 2007-04-03 16:01 & ~ & ~ & ['d2'] & ['Normal'] & ~ \\
e18 & Delivery Successful & 2007-04-03 17:59 & ~ & ~ & ['d3'] & ['Normal'] & ~ \\
e19 & Pay Order & 2007-04-06 08:00 & ['o2'] & ~ & ~ & ['Normal'] & ~ \\
e20 & Create Order & 2007-04-08 19:38 & ['o3'] & ['i7'] & ~ & ['Normal'] & ~ \\
e21 & Pick Item & 2007-04-08 23:54 & ['o3'] & ['i7'] & ~ & ['Normal'] & ~ \\
e22 & Pack Item & 2007-04-08 23:58 & ~ & ['i7'] & ['d4'] & ['Normal'] & ~ \\
e23 & Delivery Successful & 2007-04-10 11:15 & ~ & ~ & ['d4'] & ['Normal'] & ~ \\
e24 & Pay Order & 2007-04-10 12:19 & ['o3'] & ~ & ~ & ['Normal'] & ~ \\
\hline
\end{tabular}
}
\label{tab:table1}
\vspace{-6mm}
\end{table*}
Object-centric process mining is a recent branch of process mining that changes the assumption of having a unique case notion to the assumption that an event could be associated with different
objects of different object types (e.g., order, delivery). Therefore, the convergence and divergence problems can be resolved in the object-centric setting
since we do not need to associate a delivery to the order events
and an order to the delivery events.
Table \ref{tab:table1} contains a tabular representation of an object-centric event log. Each row corresponds to an event.
Each column identifies either an attribute at the event level (identifier, activity, timestamp) or the list of objects for an object type
(the orders, items, deliveries, weight classes, goods issues). What we can see in the table is the management of three orders.
Considering \emph{o1}, the order is associated to two items \emph{i1} and \emph{i2}, which are both delivered using the delivery \emph{d1}.
Despite that, in no event \emph{o1} and \emph{d1} are directly associated, so at least for orders and deliveries the convergence/divergence problem is resolved.
Object-centric event logs can be transformed (\emph{flattened}) to traditional event logs with the choice of an object type. For example, choosing the delivery as object type,
a flattened log would contain four cases (\emph{d1}, \emph{d2}, \emph{d3}, \emph{d4}). Each case would contain all the events related to the corresponding object
(for example, \emph{Pack Item}, \emph{Pack Item} and \emph{Delivery Successful} for \emph{d1}).
However, the extraction of an object-centric event log for mainstream information systems does not always solve the convergence/divergence problems,
and some strange behavior could still be contained in the event log.
We could see the following situations in the table: for the events with activity \emph{Pick Item}, the order object related to the event is unnecessarily reported (convergence problem);
all the events are associated with the same weight class \emph{Normal}.
Moreover, given the problems of scalability of most of the advanced object-centric process mining techniques (e.g., \cite{DBLP:conf/bis/LiCA17,DBLP:conf/apn/Fahland19}), keeping noisy
behavior in the event log could significantly increase the execution time of the object-centric process mining technique.
In Table \ref{tab:table1}, we have the object type \emph{Goods Issues} which is associated to a single event \emph{e12}.
Moreover, there is only an event \emph{e12} with activity \emph{Record Goods Issue}. Therefore, removing the object types \emph{Weight Classes} and \emph{Goods Issues}, which arguably are not
the most significant ones for the given event log, could lead to significantly improving the execution times of some object-centric process mining algorithms. As additional results,
the quality of the log (in the number of convergence/divergence problems) is improved by such filtering techniques.
Even with removing outlier behavior, object-centric process mining techniques could require a significant computational effort. To tackle this, along with filtering techniques,
sampling techniques might be considered. However, sampling in an object-centric event log is more challenging than in the traditional setting.
While keeping a subset of cases is enough in the traditional setting, in an object-centric setting defining a set of events to sample requires considering the interaction between
the different objects.
\begin{figure*}[!t]
\includegraphics[width=\textwidth]{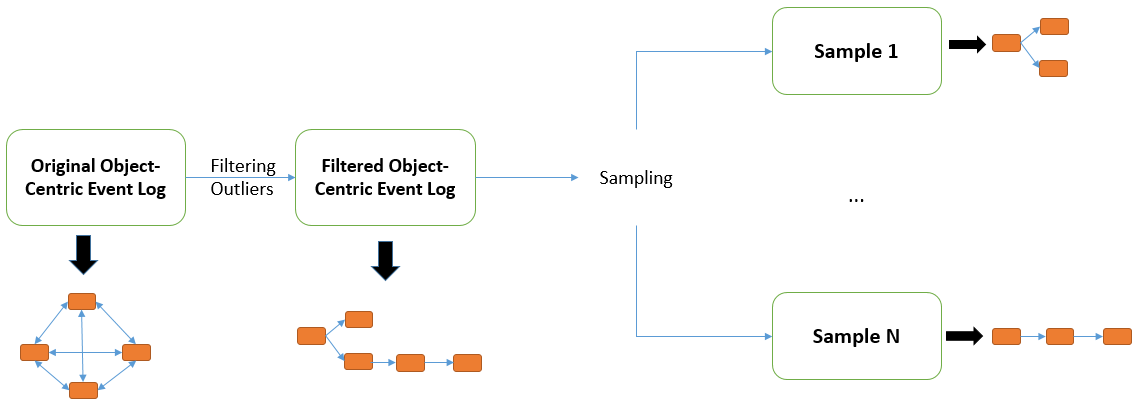}
\caption{Outline of the approach described in the paper. Object-centric process mining techniques may have scalability and quality issues starting from an object-centric event log containing lots of outlier behavior. Some filtering operations at different levels (object types, events) reduce the outlier behavior, such that
the insights are provided with higher performance and quality. Eventually, a sampling operation can split the object-centric event log into smaller units
and resolve the scalability issues.}
\label{fig:introductionOutline}
\vspace{-6mm}
\end{figure*}
In this paper, we will tackle both the filtering and sampling problems in the object-centric setting, with the purpose to enable object-centric process mining insights
on big object-centric event logs.
Several dimensions will be considered (filtering object types, events, object types per activity).
Eventually, some strategies for sampling object-centric event logs are introduced.
The rest of the paper is organized as follows. Section \ref{sec:preliminaries} introduces the core concepts in object-centric process mining.
Section \ref{sec:approach} introduces the approach (filtering and sampling object-centric event logs). Section \ref{sec:tool} presents a tool
in which the techniques are implemented. Section \ref{sec:assessment} applies the techniques to an event log extracted from an educational ERP instance.
Section \ref{sec:relatedWork} presents the related work on object-centric process mining and sampling.
Finally, Section \ref{sec:conclusion} concludes the paper.

\section{Preliminaries}
\label{sec:preliminaries}

In the following, we will introduce the definition of object-centric event log, which is the core concept on top of which the filters will be applied.
In Def. \ref{def:universesOcel}, some universes needed for the definition of OCEL are introduced.

\begin{definition}[Universes (for OCEL)]\label{def:universesOcel}
Some universes are used in the formal definition of object-centric event logs:
${U_e}$ is the universe of event identifiers; ${U_c}$ is the universe of case identifiers; ${U_{act}}$ is the universe of activities;
${U_{timest}}$ is the universe of timestamps; ${U_{att}}$ is the universe of attribute names;
${U_{val}}$ is the universe of attribute values; ${U_{typ}}$ is the universe of attribute types;
${U_{o}}$ is the universe of object identifiers; ${U_{ot}}$ is the universe of objects types.
\end{definition}

Def \ref{def:objCentrEventLog} introduces the formal definition of object-centric event log.

\begin{definition}[Object-Centric Event Log]\label{def:objCentrEventLog}
An object-centric event log is a tuple $L = (E, \allowbreak AN, \allowbreak AV, \allowbreak AT, \allowbreak OT, \allowbreak O, \allowbreak \pi_{typ}, \allowbreak \pi_{act}, \allowbreak \pi_{time}, \allowbreak \pi_{vmap}, \allowbreak \pi_{omap}, \allowbreak \pi_{otyp}, \allowbreak \pi_{ovmap}, \allowbreak \leq)$ such that:
\begin{itemize}

\item $E \subseteq {U_e}$ is the set of event identifiers.

\item $AN \subseteq {U_{att}}$ is the set of attributes names.

\item $AV \subseteq {U_{val}}$ is the set of attribute values (with the requirement that $\textrm{AN} \cap \textrm{AV} = \emptyset$).

\item $AT \subseteq {U_{typ}}$ is the set of attribute types.
 
\item $OT \subseteq {U_{ot}}$ is the set of object types.

\item $O \subseteq {U_{o}}$ is the set of object identifiers.

\item $\pi_{typ} : AN \cup AV \rightarrow AT$ is the function associating an attribute name or value to its corresponding type.

\item $\pi_{act} : E \rightarrow {U_{act}}$ is the function associating an event (identifier) to its activity.

\item $\pi_{time} : E \rightarrow {U_{timest}}$ is the function associating an event (identifier) to a timestamp.

\item $\pi_{vmap}: E \rightarrow (AN \not\rightarrow AV)$ such that
$\pi_{typ}(n) = \pi_{typ}(\pi_{vmap}(e)(n)) \quad \allowbreak \forall  e \in E ~ \allowbreak \forall n \in \textrm{dom}(\pi_{vmap}(e))$
is the function associating an event (identifier) to its attribute value assignments.

\item $\pi_{omap} : E \rightarrow \mathcal{P}(O)$ is the function associating an event (identifier) to a set of related object identifiers.

\item $\pi_{otyp} \in O \rightarrow OT$ assigns precisely one object type to each object identifier.

\item $\pi_{ovmap} : O \rightarrow (AN \not\rightarrow AV)$ such that
$\pi_{typ}(n) = \pi_{typ}(\pi_{ovmap}(o)(n)) \quad \allowbreak \forall n \in \textrm{dom}(\pi_{ovmap}(o)) ~ \allowbreak \forall o \in O$
is the function associating an object to its attribute value assignments.

\item $\leq$ is a total order (i.e., it respects the antisymmetry, transitivity, and connexity properties).

\end{itemize}
\end{definition}

Recently, the OCEL standard \url{http://www.ocel-standard.org/}
has been introduced for the storage of object-centric event logs \cite{DBLP:conf/adbis/GhahfarokhiPBA21}. Two different implementations of the standard are provided
(JSON-OCEL and XML-OCEL), which are based on popular file formats.

\section{Approach}
\label{sec:approach}

The approach section is organized as follows. First, the different filters on the outliers (object types, events, object types per activity)
are introduced. Then, the sampling methodologies are introduced.

\subsection{Filter Outlier Object Types}
\label{subsec:filterOutlierObjectTypes}

Starting from an object-centric event log, it's possible to filter a subset of object types. This will reduce the number of possible dimensions of the object-centric event log
on top of which the interactions between the different object types are considered and possibly resolve some divergence problems.
Def. \ref{def:filteringObjectTypes} introduces object type-based filtering.

\begin{definition}[Filtering a Subset of Object Types]\label{def:filteringObjectTypes}
Let
$L = (E, \allowbreak AN, \allowbreak AV, \allowbreak AT, \allowbreak OT, \allowbreak O, \allowbreak \pi_{typ}, \allowbreak \pi_{act}, \allowbreak \pi_{time}, \allowbreak \pi_{vmap}, \allowbreak \pi_{omap}, \allowbreak \pi_{otyp}, \allowbreak \pi_{ovmap}, \allowbreak \leq)$
be an object-centric event log. Given a set of object types $\bar{OT} \subseteq OT$, we define $\bar{O} = \{ o \in O ~ \arrowvert ~ \pi_{otyp}(o) \in \bar{OT} \}$ as the set of objects having a type included in $\bar{OT}$.
We define the filtered log
$L_{OT \arrowvert \bar{OT}} = (E, \allowbreak AN, \allowbreak AV, \allowbreak AT, \allowbreak \bar{OT}, \allowbreak \bar{O}, \allowbreak \pi_{typ}, \allowbreak \pi_{act}, \allowbreak \pi_{time}, \allowbreak \pi_{vmap}, \allowbreak \pi_{omap}', \allowbreak \restr{\pi_{otyp}}{\bar{O}}, \allowbreak \restr{\pi_{ovmap}}{\bar{O}}, \allowbreak \leq)$
where $\pi_{omap}'(e) = \{ o \in \pi_{omap}(e) ~ \arrowvert ~ o \in \bar{O} \}$.
\end{definition}

In Def. \ref{def:outlierFiltersObjectTypes}, some filters on the object types are introduced. The first two filters (OTF1 and OTF2) count the number of different objects of an object type,
the number of different events related to objects of a given type, and keep only the ones that have frequency $\geq n$. The third (OTF3) measures the divergence problems
for an object type since it counts if the activities are often repeated in the lifecycle of the objects of a type.

\begin{definition}[Outlier Filters on Object Types]\label{def:outlierFiltersObjectTypes}
Let
$L = (E, \allowbreak AN, \allowbreak AV, \allowbreak AT, \allowbreak OT, \allowbreak O, \allowbreak \pi_{typ}, \allowbreak \pi_{act}, \allowbreak \pi_{time}, \allowbreak \pi_{vmap}, \allowbreak \pi_{omap}, \allowbreak \pi_{otyp}, \allowbreak \pi_{ovmap}, \allowbreak \leq)$
be an object-centric event log. We define the following filters at the object type level:
\begin{enumerate}[label=OTF\arabic*]
\item (\emph{Minimum Number of Different Objects per Object Type}) given $n \in \mathbb{N}$, we define:
$\bar{OT} = \{ ot \in OT ~ \arrowvert ~ \arrowvert \{ o \in O ~ \arrowvert ~ \pi_{otyp}(o) = ot \} \arrowvert \geq n \}$
\item (\emph{Minimum Number of Events Associated per Object Type}) given $n \in \mathbb{N}$, we define:
$\bar{OT} = \{ ot \in OT ~ \arrowvert ~ \arrowvert \{ e \in E ~ \arrowvert ~ \exists_{o \in \pi_{omap}(e)} \pi_{otyp}(o) = ot \} \arrowvert \geq n \}$
\item (\emph{Minimum Ratio of Unique Activities}) given $r \in [0, 1]$, we define:
$$\bar{OT} = \{ ot \in OT ~ \arrowvert ~ \frac{\sum_{o \in O, \pi_{otyp}(o) = ot} \arrowvert \{ \pi_{act}(e) ~ \arrowvert ~ e \in E \wedge o \in \pi_{omap}(e) \} \arrowvert}{\sum_{o \in O, \pi_{otyp}(o) = ot} \arrowvert \{ e \in E ~ \arrowvert ~ o \in \pi_{omap}(e) \} \arrowvert } \geq r \} $$
\end{enumerate}
\end{definition}

Table \ref{tab:table2} can be obtained applying filtering on object types. The object type \emph{Goods Issues} can be filtered out by OTF1 and OTF2
(since there is only an event associated with objects of the given type, and a single object is associated with the given type).
The object type \emph{Weight Classes} can be filtered out either by OTF1 (since there is a single object in the given object type) and OTF3
(for the object \emph{Normal}, there are only six unique activities out of 24, so any ratio above $0.25$ will remove the object type).

\subsection{Filter Outlier Events}
\label{subsec:filterOutlierEvents}

\begin{table*}[!b]
\vspace{-7mm}
\caption{Filtered object-centric event log, using the filters at the object type level, starting from the log introduced in Table \ref{tab:table1}. The activities highlighted in bold
are essential events (Def. \ref{def:essentialEvents}).
The objects colored in gray belong to harmful object types-activities associations (Def. \ref{def:outlierFiltersObjectTypesActivitiesRelations}).}
\centering
\resizebox{0.75\textwidth}{!}{ 
\begin{tabular}{|lcc|ccc|}
\hline
{\bf Event ID} & {\bf Event Activity} & {\bf Event Timestamp} & {\bf Orders} & {\bf Items} & {\bf Deliveries} \\
\hline
e1 & {\bf Create Order} & 2007-04-01 07:29 & ['o1'] & ['i1', 'i2'] & ~ \\
e2 & Pick Item & 2007-04-01 15:36 & \textcolor{gray}{['o1']} & ['i1'] & ~ \\
e3 & {\bf Pack Item} & 2007-04-01 22:01 & ~ & ['i1'] & ['d1'] \\
e4 & Pick Item & 2007-04-01 22:15 & \textcolor{gray}{['o1']} & ['i2'] & ~ \\
e5 & {\bf Pack Item} & 2007-04-01 23:19 & ~ & ['i2'] & ['d1']  \\
e6 & {\bf Deliver Successful} & 2007-04-02 10:01 & ~ & ~ & ['d1']  \\
e7 & {\bf Pay Order} & 2007-04-02 11:15 & ['o1'] & ~  & ~  \\
e8 & {\bf Create Order} & 2007-04-02 12:15 & ['o2'] & ['i4', 'i5', 'i6'] & ~ \\
e9 & Pick Item & 2007-04-03 02:19 & \textcolor{gray}{['o2']} & ['i4'] & ~ \\
e10 & {\bf Pack Item} & 2007-04-03 03:17 & ~ & ['i4'] & ['d2'] \\
e11 & Pick Item & 2007-04-03 05:02 & \textcolor{gray}{['o2']} & ['i5'] & ~ \\
e12 & Record Goods Issue & 2007-04-03 05:07 & ~ & ['i5'] & ~ \\
e13 & Pick Item & 2007-04-03 05:15 & \textcolor{gray}{['o2']} & ['i5'] & ~ \\
e14 & {\bf Pack Item} & 2007-04-03 05:21 & ~ & ['i5'] & ['d3'] \\
e15 & Pick Item & 2007-04-03 05:36 & \textcolor{gray}{['o2']} & ['i6'] & ~\\
e16 & {\bf Pack Item} & 2007-04-03 05:37 & ~ & ['i6'] & ['d3'] \\
e17 & {\bf Delivery Successful} & 2007-04-03 16:01 & ~ & ~ & ['d2'] \\
e18 & {\bf Delivery Successful} & 2007-04-03 17:59 & ~ & ~ & ['d3'] \\
e19 & {\bf Pay Order} & 2007-04-06 08:00 & ['o2'] & ~ & ~ \\
e20 & {\bf Create Order} & 2007-04-08 19:38 & ['o3'] & ['i7'] & ~ \\
e21 & Pick Item & 2007-04-08 23:54 & \textcolor{gray}{['o3']} & ['i7'] & ~ \\
e22 & {\bf Pack Item} & 2007-04-08 23:58 & ~ & ['i7'] & ['d4'] \\
e23 & {\bf Delivery Successful} & 2007-04-10 11:15 & ~ & ~ & ['d4'] \\
e24 & {\bf Pay Order} & 2007-04-10 12:19 & ['o3'] & ~ & ~ \\
\hline
\end{tabular}
}
\label{tab:table2}
\end{table*}

Another type of filtering possible on object-centric event logs is event-based filtering. Essentially, a subset of events of the original object-centric event log is selected.

\begin{definition}[Filtering a Subset of Events]\label{def:filteringEvents}
Let
$L = (E, \allowbreak AN, \allowbreak AV, \allowbreak AT, \allowbreak OT, \allowbreak O, \allowbreak \pi_{typ}, \allowbreak \pi_{act}, \allowbreak \pi_{time}, \allowbreak \pi_{vmap}, \allowbreak \pi_{omap}, \allowbreak \pi_{otyp}, \allowbreak \pi_{ovmap}, \allowbreak \leq)$
be an object-centric event log. Given a set of events $\bar{E} \subseteq E$, we define the filtered log
$L_{E \arrowvert \bar{E}} = (\bar{E}, \allowbreak AN, \allowbreak AV, \allowbreak AT, \allowbreak OT, \allowbreak O, \allowbreak \restr{\pi_{typ}}{\bar{E}}, \allowbreak \restr{\pi_{act}}{\bar{E}}, \allowbreak \restr{\pi_{time}}{\bar{E}}, \allowbreak \restr{\pi_{vmap}}{\bar{E}}, \allowbreak \restr{\pi_{omap}}{\bar{E}}, \allowbreak \pi_{otyp}, \allowbreak \pi_{ovmap}, \allowbreak \restr{\leq}{\bar{E} \times \bar{E}})$
\end{definition}

In the following, some criteria will be introduced to grade the importance of events. This will help to identify the events to keep/remove in the filtering.
First, in Def. \ref{def:eventsLifecycleInteraction} the concepts of \emph{lifecycle} (the sequence of events related to an object)
and \emph{interaction lifecycle} (the sequence of events related to a couple of objects; so, the steps in which the interactions between the two objects occur)
are introduced.

\begin{definition}[Objects Lifecycle and Interaction Lifecycle]\label{def:eventsLifecycleInteraction}
Let
$L = (E, \allowbreak AN, \allowbreak AV, \allowbreak AT, \allowbreak OT, \allowbreak O, \allowbreak \pi_{typ}, \allowbreak \pi_{act}, \allowbreak \pi_{time}, \allowbreak \pi_{vmap}, \allowbreak \pi_{omap}, \allowbreak \pi_{otyp}, \allowbreak \pi_{ovmap}, \allowbreak \leq)$
be an object-centric event log. Given an object $o \in O$, we define its lifecycle as the ordered sequence of events
$\textrm{lif}(o) = [e_1, \ldots, e_n]$
such that $o \in \pi_{omap}(e_i)$ for $i \in \{ 1, \ldots, n \}$ and $e_{i-1} \leq e_{i}$ for $i \in \{ 2, \ldots, n \}$.
We define the interaction lifecycle between two objects $o_1, o_2 \in O$ as the ordered sequence of events
$\textrm{intlif}(o_1, o_2) = [e_1, \ldots, e_m]$
where $o_1 \in \pi_{omap}(e_i) \wedge o_2 \in \pi_{omap}(e_i)$ for $i \in \{ 1, \ldots, m \}$ and $e_{i-1} \leq e_{i}$ for $i \in \{ 2, \ldots, m \}$.
\end{definition}

\emph{Essential events} are defined in Def. \ref{def:essentialEvents} as the ones which either start/complete the lifecycle of an object, start/complete an interaction between two objects,
or show a synchronization of the lifecycle \cite{DBLP:conf/apn/Fahland19}.

\begin{definition}[Essential Events]\label{def:essentialEvents}
Let
$L = (E, \allowbreak AN, \allowbreak AV, \allowbreak AT, \allowbreak OT, \allowbreak O, \allowbreak \pi_{typ}, \allowbreak \pi_{act}, \allowbreak \pi_{time}, \allowbreak \pi_{vmap}, \allowbreak \pi_{omap}, \allowbreak \pi_{otyp}, \allowbreak \pi_{ovmap}, \allowbreak \leq)$
be an object-centric event log. We define an event $e \in E$ as \emph{essential} if one of the following conditions is met:
\begin{enumerate}[label=EE\arabic*]
\item (\emph{Starting the Lifecycle of an Object}) $\exists_{o \in O} ~ e = \textrm{lif}(o)(1)$.
\item (\emph{Ending the Lifecycle of an Object}) $\exists_{o \in O} ~ e = \textrm{lif}(o)(\arrowvert \textrm{lif}(o) \arrowvert)$.
\item (\emph{Starting the Interaction between two Objects}) $\exists_{o_1, o_2 \in O, o_1 \neq o_2} ~ e = \textrm{intlif}(o_1, o_2)(1)$.
\item (\emph{Ending the Interaction between two Objects})
$\exists_{o_1, o_2 \in O, o_1 \neq o_2} ~ \allowbreak e = \textrm{intlif}(o_1, o_2)(\allowbreak \arrowvert \textrm{intlif}(o_1, o_2) \arrowvert)$
\item (\emph{Synchronization between the Lifecycles})
$$\exists_{o_1, o_2 \in O, o_1 \neq o_2, e_1 \leq e_2 \leq e_3 \in E} ~ e_1, e_3 \in \textrm{intlif}(o_1, o_2) ~ \wedge ~ e_2 \not\in \textrm{intlif}(o_1, o_2) ~ \wedge ~ e_2 \in \textrm{lif}(o_1) \cup \textrm{lif}(o_2)$$
\end{enumerate}
We define $\textrm{essev}(L)$ as the set of essential events of the object-centric event log.
\end{definition}

Essential events, together with the frequency of activities, can be used to formulate reasonable filters at the event level.
We can either consider each filter in isolation or consider the combination between filtering essential events and frequency of activities:
an event is kept in the log if it is essential or its number of occurrences exceeds a threshold.

\begin{definition}[Outlier Filters on Events]\label{def:outlierFiltersEvents}
Let
$L = (E, \allowbreak AN, \allowbreak AV, \allowbreak AT, \allowbreak OT, \allowbreak O, \allowbreak \pi_{typ}, \allowbreak \pi_{act}, \allowbreak \pi_{time}, \allowbreak \pi_{vmap}, \allowbreak \pi_{omap}, \allowbreak \pi_{otyp}, \allowbreak \pi_{ovmap}, \allowbreak \leq)$
be an object-centric event log. We define the following filters at the event level:
\begin{enumerate}[label=OE\arabic*]
\item (\emph{Filter Activities Occurrences}) Let $E_a = \{ e \in E ~ \arrowvert ~ \pi_{act}(e) = a \}$. Given $n \in \mathbb{N}$, we define
$\bar{E} = \{ e \in E ~ \arrowvert ~ \arrowvert E_{\pi_{act}(e)} \arrowvert \geq n \}$
\item (\emph{Filter on Essential Events}) $\bar{E} = \textrm{essev}(L)$.
\item (\emph{Filter Activities Occurrences or Essential Events}) Let $E_a = \{ e \in E ~ \arrowvert ~ \pi_{act}(e) = a \}$. Given $n \in \mathbb{N}$, we define:
$\bar{E} = \{ e \in E ~ \arrowvert ~ \arrowvert E_{\pi_{act}(e)} \arrowvert \geq n ~ \vee ~ e \in \textrm{essev}(L) \}$
\end{enumerate}
\end{definition}

Let's consider the event log reported in Table \ref{tab:table2}. The events highlighted in bold are essential events according to one of the criteria defined in Def. \ref{def:essentialEvents}.

\subsection{Filter Object Types per Activity}
\label{subsec:filterObjectTypesPerActivity}

Another possibility is to filter the related objects to an event based on an activity-object type association. An object type, such as \emph{Orders} in Table \ref{tab:table2},
is important for the definition of events having some activities (such as \emph{Create Order} and \emph{Pay Order}) but might lead to divergence issues for events having some other activities
(such as \emph{Pick Item}). Therefore, a strategy is to leave the associations between the objects of type \emph{Order} and the events having as activity \emph{Create Order} or \emph{Pay Order},
but remove the associations with the events having as activity \emph{Pick Item}. Def. \ref{def:filteringOtActivitiesAssociations} introduces such filtering.

\begin{definition}[Filtering Object Types-Activities Relations]\label{def:filteringOtActivitiesAssociations}
Let
$L = (E, \allowbreak AN, \allowbreak AV, \allowbreak AT, \allowbreak OT, \allowbreak O, \allowbreak \pi_{typ}, \allowbreak \pi_{act}, \allowbreak \pi_{time}, \allowbreak \pi_{vmap}, \allowbreak \pi_{omap}, \allowbreak \pi_{otyp}, \allowbreak \pi_{ovmap}, \allowbreak \leq)$
be an object-centric event log. Let $\bar{OA} \subseteq OT \times \{ \pi_{act}(e) ~ \arrowvert ~ e \in E \}$ be a set of allowed object types - activities relations.
We consider the filtered event log $L_{OA \arrowvert \bar{OA}} = (E, \allowbreak AN, \allowbreak AV, \allowbreak AT, \allowbreak OT, \allowbreak O, \allowbreak \pi_{typ}, \allowbreak \pi_{act}, \allowbreak \pi_{time}, \allowbreak \pi_{vmap}, \allowbreak \pi_{omap}', \allowbreak \pi_{otyp}, \allowbreak \pi_{ovmap}, \allowbreak \leq)$
where
$$\pi_{omap}'(e) = \{ o \in \pi_{omap}(e) ~ \arrowvert ~ (\pi_{otyp}(o), \pi_{act}(e)) \in \bar{OA} \}$$
\end{definition}

A possible strategy to detect the associations between object types and activities is the one described in Def. \ref{def:outlierFiltersObjectTypesActivitiesRelations}.
Essentially, for every possible combination of object types and activities, the number of objects of the given object type associated with events having the given activity is counted.
This is divided by the summation, for all the objects of the given object type, of the number of events having the given activity related to the given object.
A number between 0 and 1 is obtained with this division. The number is lower when several events of the given activity are related to the same object. Therefore, the associations
leading to a divergence problem could be pruned out using this number.

\begin{definition}[Outlier Filter on Object Types-Activities Relations]\label{def:outlierFiltersObjectTypesActivitiesRelations}
Let
$L = (E, \allowbreak AN, \allowbreak AV, \allowbreak AT, \allowbreak OT, \allowbreak O, \allowbreak \pi_{typ}, \allowbreak \pi_{act}, \allowbreak \pi_{time}, \allowbreak \pi_{vmap}, \allowbreak \pi_{omap}, \allowbreak \pi_{otyp}, \allowbreak \pi_{ovmap}, \allowbreak \leq)$
be an object-centric event log. Given $r \in [0, 1]$, we define the following filter (ratio of unique objects per type per activity):
\begin{align*}
\bar{OA} = \{ (ot, a) \in OT \times \{ \pi_{act}(e) ~ \arrowvert ~ e \in E \} ~ \arrowvert \\
\frac{\sum_{o \in O, \pi_{otyp}(o) = ot} \mathbbm{1}_{\exists e \in E, \pi_{act}(e) = a ~ \wedge ~ o \in \pi_{omap}(e)}}{\sum_{o \in O, \pi_{otyp}(o) = ot} \arrowvert \{ e \in E ~ \arrowvert ~ \pi_{act}(e) = a ~ \wedge ~ o \in \pi_{omap}(e) \} \arrowvert } \geq r \}
\end{align*}
\end{definition}

\begin{figure}[ht]
\vspace{-5mm}
\centering
\includegraphics[width=\textwidth]{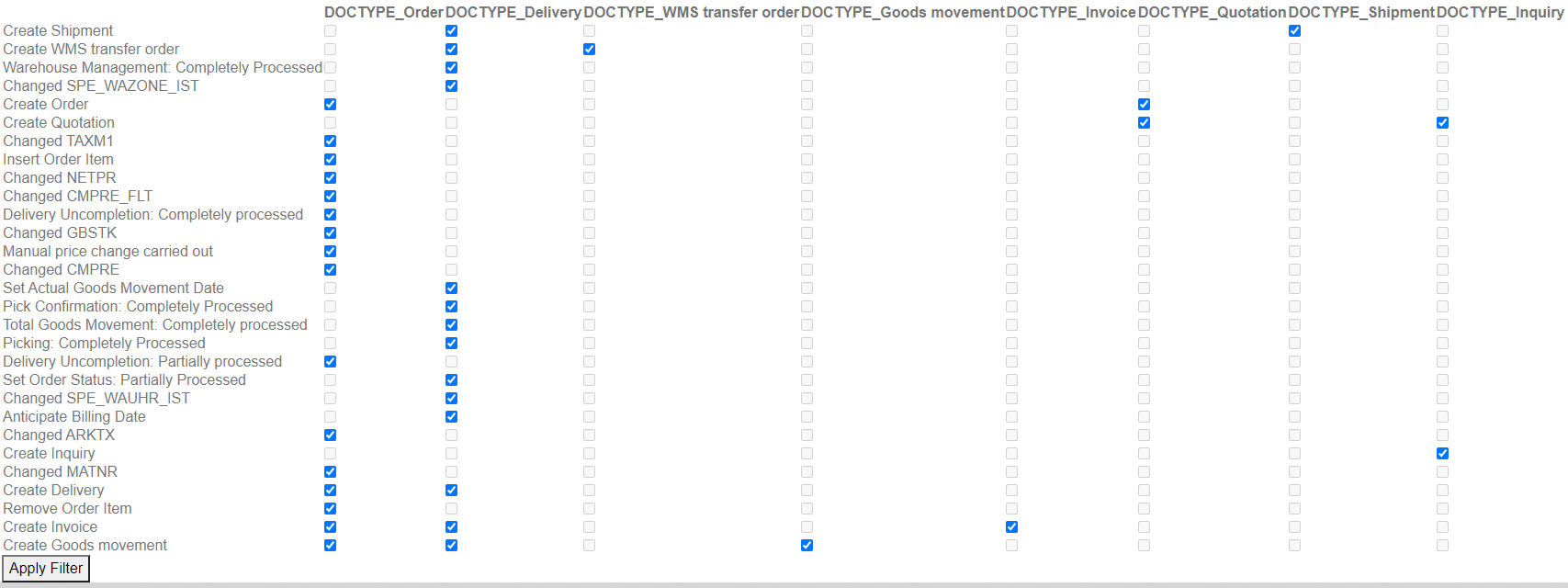}
\caption{Manual selection of the activity-object type relations.}
\label{fig:manualSelectionActivityObjectType}
\vspace{-5mm}
\end{figure}

In Table \ref{tab:table2}, the objects of type \emph{Orders} related to events having activity \emph{Pick Item} are highlighted because the association leads to divergence problem.
This is identified using the definition Def. \ref{def:outlierFiltersObjectTypesActivitiesRelations}.
An alternative is a manual selection of the association between activities and object types (see Fig. \ref{fig:manualSelectionActivityObjectType}).

\subsection{Sampling}
\label{subsec:sampling}

In this section, we will describe some sampling strategies that could be used on object-centric event logs. The first three are random sampling on different dimensions (events, objects, object types).
The fourth considers as samples the connected events of the log. Initially, Def. \ref{def:filteringObjects} is introduced to allow filtering on objects.

\begin{definition}[Filtering a Subset of Objects]\label{def:filteringObjects}
Let
$L = (E, \allowbreak AN, \allowbreak AV, \allowbreak AT, \allowbreak OT, \allowbreak O, \allowbreak \pi_{typ}, \allowbreak \pi_{act}, \allowbreak \pi_{time}, \allowbreak \pi_{vmap}, \allowbreak \pi_{omap}, \allowbreak \pi_{otyp}, \allowbreak \pi_{ovmap}, \allowbreak \leq)$
be an object-centric event log. Given a set of objects $\bar{O} \subseteq O$,
we define the filtered log
$L_{O \arrowvert \bar{O}} = (E, \allowbreak AN, \allowbreak AV, \allowbreak AT, \allowbreak OT, \allowbreak \bar{O}, \allowbreak \pi_{typ}, \allowbreak \pi_{act}, \allowbreak \pi_{time}, \allowbreak \pi_{vmap}, \allowbreak \pi_{omap}', \allowbreak \restr{\pi_{otyp}}{\bar{O}}, \allowbreak \restr{\pi_{ovmap}}{\bar{O}}, \allowbreak \leq)$
where $\pi_{omap}'(e) = \{ o \in \pi_{omap}(e) ~ \arrowvert ~ o \in \bar{O} \}$.
\end{definition}

Def. \ref{def:samplingStrategies} introduces the sampling strategies formally.

\begin{definition}[Sampling Strategies]\label{def:samplingStrategies}
Let
$L = (E, \allowbreak AN, \allowbreak AV, \allowbreak AT, \allowbreak OT, \allowbreak O, \allowbreak \pi_{typ}, \allowbreak \pi_{act}, \allowbreak \pi_{time}, \allowbreak \pi_{vmap}, \allowbreak \pi_{omap}, \allowbreak \pi_{otyp}, \allowbreak \pi_{ovmap}, \allowbreak \leq)$
be an object-centric event log. We can define the following sampling strategies:
\begin{enumerate}[label=SS\arabic*]
\item (Random sampling on events) A random subset of events $\bar{E} \subseteq E$ is chosen. The sampled log is $L_{E \arrowvert \bar{E}}$ (Def. \ref{def:filteringEvents}).
\item (Random sampling on objects) A random subset of events $\bar{O} \subseteq O$ is chosen. The sampled log is $L_{O \arrowvert \bar{O}}$ (Def. \ref{def:filteringObjects}).
\item (Random sampling on object types) A random subset of object types $\bar{OT} \subseteq OT$ is chosen. The sampled log is $L_{OT \arrowvert \bar{OT}}$ (Def. \ref{def:filteringObjectTypes}).
\item (Sampling connected events) Given the relation $R \subseteq E \times E$, $(e_1, e_2) \in R \iff \pi_{omap}(e_1) \cap \pi_{omap}(e_2) \neq \emptyset$,
consider the transitive closure\footnote{$R^{+} = \cup_{i=1}^{\infty} R^i$ where $R^i$ is defined inductively as $R^1 = R$ and for $i > 0$ $R^{i+1} = R \circ R^i$.} $R^{+}$ of $R$.
The equivalence classes of $R^{+}$ are the samples.
\end{enumerate}
\end{definition}

Random sampling on object types removes some object types from the log, but the insights on the remaining object types are entirely valid.
Random sampling events/objects trim the lifecycle of objects and remove relationships between them, so the quality of insights obtained from
object-centric process mining techniques is hampered. The fourth strategy divides the event log into ``self-contained units'' on top of which meaningful insights can be discovered.

\section{Tool}
\label{sec:tool}

\begin{figure}[ht]
\vspace{-5mm}
\centering
\includegraphics[width=\textwidth]{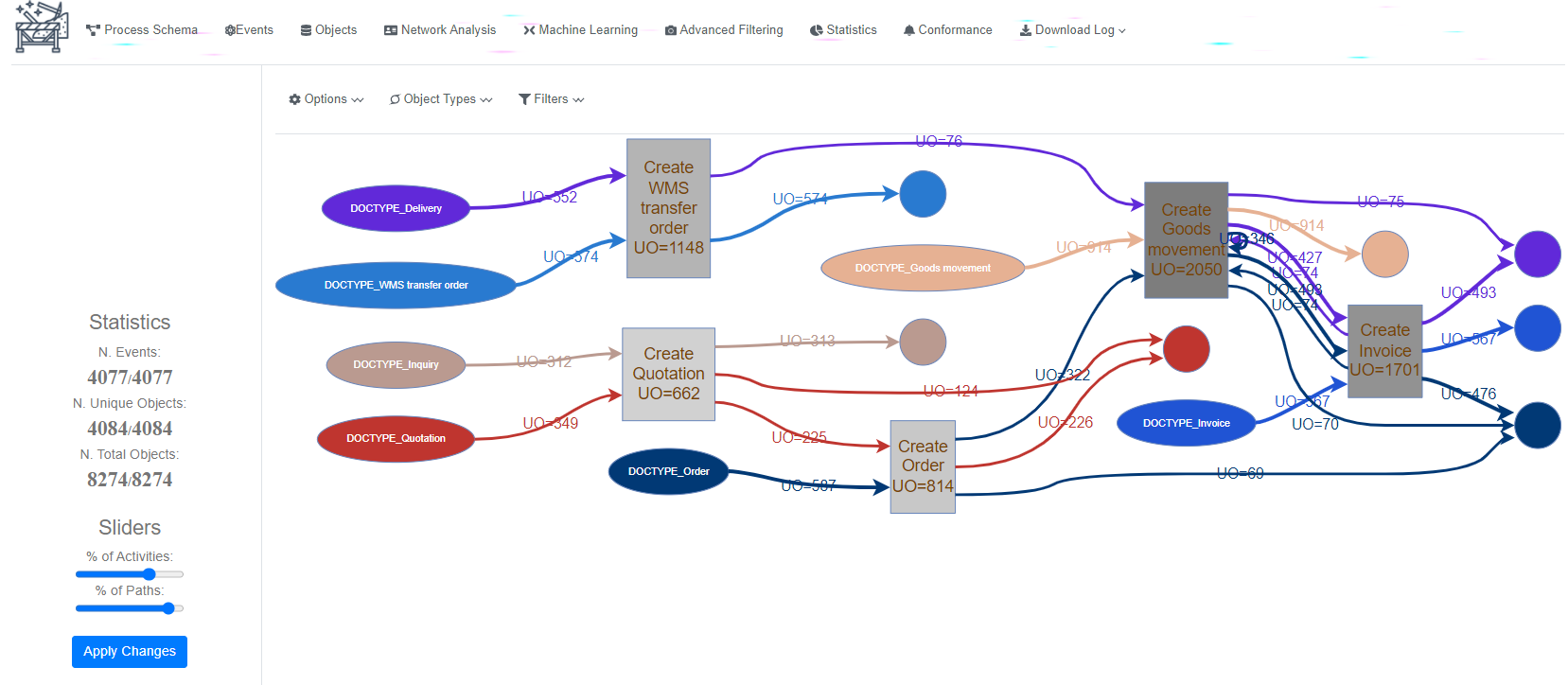}
\caption{Screenshot of the process schema component of the OCPM tool.}
\label{fig:ocpmTool}
\vspace{-5mm}
\end{figure}

This section presents OCPM, a web interface for object-centric process mining. The repository of the process is \\ \url{https://github.com/Javert899/pm4js-sandbox}.
The application can be run by either 1) cloning the Github repository and 2) launching the file \emph{ocel.html}, or accessing the application deployed at
\url{http://www.ocel-standard.org/ocpm/ocel.html}. The application is made by several components (process schema, events explorer, objects explorer, network analysis,
filtering, statistics, conformance). The filtering component could be accessed by clicking ``Advanced Filtering'' on the top. There, the filters proposed in this paper
are proposed. In addition, the deployed application is pre-loaded with the event log used in the assessment. Fig. \ref{fig:ocpmTool} shows the process schema component of the tool.

\section{Assessment}
\label{sec:assessment}

The assessment proposed in this section is based on the tool OCPM (presented in the previous section) and an Order-to-Cash log extracted from an educational SAP instance.
The original event log is composed by $4077$ events, $4084$ objects and $8$ different object types (\emph{Inquiry} with $316$ objects, \emph{Quotation} with $361$ objects,
\emph{Order} with $640$ objects, \emph{Delivery} with $614$ objects, \emph{Goods Movement} with $914$ objects, \emph{Shipment} with $96$ objects, \emph{WMS transfer order} with $576$ objects, and \emph{Invoice} with $567$ objects).

\emph{Filter on Object Types}: at first, we would like to filter out some of the less significant object types. If we apply the filter OTF2 (filter object types having a minimum number of related events) with $n = 700$,
the object types of the event log are reduced to \emph{Order}, \emph{Delivery} and \emph{Goods Movement} (the object types with the most frequent occurrence in the log).
The event log is reduced to $2168$ objects ($53 \%$).

\emph{Filter on Events}: at this point, we would like to filter the object-centric event log keeping only the essential events. In an Order-to-Cash process, the \emph{change activities}
operate on an existing object, and we could remove all the events which do not start/terminate the lifecycle of an object or start/terminate/synchronize the interaction between
two objects. Therefore, the event log is further reduced to $2686$ events ($65 \%$).

\emph{Sampling}: using the sampling strategy \emph{SS4} in Def. \ref{def:samplingStrategies}, we get $657$ samples, with the most significant counting $5$ events.
Therefore, the event log is split into many small sub-logs containing only parts of the behavior.

\section{Related Work}
\label{sec:relatedWork}

This section presents the related work on object-centric process mining and filtering/sampling event logs.

\emph{Object-centric Process Mining}: in \cite{DBLP:conf/sefm/Aalst19}, two of the main problems of traditional event logs (convergence/divergence) are presented as motivation
to work with object-centric process mining. Different process models have been proposed, including \cite{DBLP:conf/bis/LiCA17} (object-centric behavioral constraints models;
from an object-centric event log, a declarative model is discovered but with limited scalability due to the computationally-expensive constraints), \cite{DBLP:conf/simpda/BertiA19}
(a model is discovered as collation of directly-follows graphs) and \cite{DBLP:journals/fuin/AalstB20} (discovering object-centric Petri nets). In \cite{DBLP:conf/apn/Fahland19},
unbounded synchronizations in processes with many-to-many interactions are described, which require to know all the synchronizations between the lifecycles of the object. The OCEL standard for the storage of object-centric event logs is proposed in
\cite{DBLP:conf/adbis/GhahfarokhiPBA21}.

\emph{Filtering and Sampling}: many scientific contributions exist in this area. For filtering, we cite as references \cite{DBLP:journals/tkde/ConfortiRH17} 
and \cite{DBLP:conf/bpm/SaniZA17}. For sampling, we cite \cite{DBLP:conf/icpm/KabierskiNGW21}.

\section{Conclusion}
\label{sec:conclusion}

In this paper, a methodology to filter outliers and sample object-centric event log is presented. Several dimensions (events, objects, object types) are considered for such operations.
Filtering is done to increase the quality of the object-centric event log (for example, reducing the divergence problems). This helps to reduce the complexity of
object-centric process mining operations and the resulting insights. A sampling operation can be applied to the original/filtered object-centric event log.
This is especially useful on real-life datasets since some object-centric process mining techniques have limited scalability.
We also propose a web-based tool implementing the aforementioned filtering and sampling operations. The assessment on an O2C object-centric event log extracted from an educational instance
shows that the filtering technique helps to reduce the infrequent behavior in the object-centric event log, and the sampling techniques return several samples out of the considered log.

\emph{Future Work}: The set of techniques proposed in the current paper is based on simple metrics. More advanced metrics could be considered. For example, having a set of features
associated to each object (such as metrics on the lifecycle, on the interaction between objects, and on the object attributes) enables the application of anomaly detection techniques
(such as isolation forests).

\section*{Acknowledgements}

We thank the Alexander von Humboldt (AvH) Stiftung for supporting our research.

\bibliographystyle{splncs04}
\bibliography{sampling_ocel}

\end{document}